\documentclass{aa}
\usepackage[utf8]{inputenc}
\usepackage[varg]{txfonts}
\usepackage[colorlinks,citecolor=blue,linkcolor=blue]{hyperref}
\usepackage{amsmath}
\usepackage{graphicx}
\usepackage[english]{babel}
\usepackage{siunitx}
\usepackage{float}
\usepackage{url}
\usepackage{longtable}
\usepackage{fancyhdr}
\usepackage{natbib}
\usepackage[normalem]{ulem}

\makeatletter
\def\instrefs#1{{\def\scsep{\def\scsep{,}}\@for\w:=#1\do{\scsep\ref{inst:\w}}}}
\renewcommand{\inst}[1]{\unskip$^{\instrefs{#1}}$}

\renewcommand*\aa@pageof{, page \thepage{} of \pageref*{LastPage}} 

\makeatother

\title{Precise radial velocities of giant stars}
\subtitle{XIII. A second Jupiter orbiting in 4:3 resonance in the 7~CMa system
\thanks{
Based on observations collected at Lick Observatory, University of California.} 
\fnmsep 
\thanks{
Based on observations collected at the European Organization for Astronomical Research in the Southern Hemisphere under ESO programmes 
078.C-0751,
079.C-0657,
081.C-0802,
082.C-0427,
289.C-5053,
0100.C-0414 and
0101.C-0232} }

\author{R.~Luque\inst{iac,ull,lsw} 
     \and
        T.~Trifonov\inst{mpia}
     \and
        S.~Reffert\inst{lsw}
     \and
        A.~Quirrenbach\inst{lsw}
     \and
        M.~H.~Lee\inst{hongkong1,hongkong2}
     \and
        S.~Albrecht\inst{aarhus}
     \and
        M.~Fredslund~Andersen\inst{aarhus}
     \and
        V.~Antoci\inst{aarhus}
     \and
        F.~Grundahl\inst{aarhus}
     \and
        C.~Schwab\inst{mq}  
     \and
        V.~Wolthoff\inst{lsw}  
        }

\institute{
\label{inst:iac}Instituto de Astrof\'isica de Canarias (IAC), 38205 La Laguna, Tenerife, Spain; \email{rluque@iac.es}
\and 
\label{inst:ull}Departamento de Astrof\'isica, Universidad de La Laguna (ULL), 38206, La Laguna, Tenerife, Spain
\and
\label{inst:lsw}Landessternwarte, Zentrum f\"ur Astronomie der Universit\"at Heidelberg, K\"onigstuhl 12, 69117 Heidelberg, Germany
\and 
\label{inst:mpia}Max-Planck-Institut f\"ur Astronomie, K\"onigstuhl 17, 69117 Heidelberg, Germany
\and
\label{inst:hongkong1}Department of Earth Sciences, The University of Hong Kong, Pokfulam Road, Hong Kong
\and
\label{inst:hongkong2}Department of Physics, The University of Hong Kong, Pokfulman Road, Hong Kong
\and
\label{inst:aarhus}Stellar Astrophysics Centre, Department of Physics and Astronomy, Aarhus University, Ny Munkegade 120, DK-8000 Aarhus C, Denmark
\and
\label{inst:mq}Department of Physics and Astronomy, Macquarie University, North Ryde, NSW 2109, Australia
           }
           
\date{}

\abstract{
We report the discovery of a second planet orbiting the K giant star 7~CMa based on 166 high-precision radial velocities obtained with Lick, HARPS, UCLES and SONG. The periodogram analysis reveals two periodic signals of approximately 745 and 980\,d, associated to planetary companions. A double-Keplerian orbital fit of the data reveals two Jupiter-like planets with minimum masses $m_b\sin i  \sim 1.9 \,\mathrm{M_{J}}$ and $m_c\sin i \sim 0.9 \,\mathrm{M_{J}}$, orbiting at semi-major axes of $a_b \sim 1.75\,\mathrm{au}$ and $a_c \sim 2.15\,\mathrm{au}$, respectively. Given the small orbital separation and the large minimum masses of the planets close encounters may occur within the time baseline of the observations, thus, a more accurate N-body dynamical modeling of the available data is performed. The dynamical best-fit solution leads to collision of the planets and we explore the long-term stable configuration of the system in a Bayesian framework, confirming that 13\% of the posterior samples are stable for at least 10\,Myr. The result from the stability analysis indicates that the two-planets are trapped in a low-eccentricity 4:3 mean-motion resonance. This is only the third discovered system to be inside a 4:3 resonance, making it very valuable for planet formation and orbital evolution models.
}


\begin{document}

\maketitle

\section{Introduction}

Today, about 4000 exoplanets around about roughly 3000 host stars are known. Most have been found with the transiting method, while 529 systems so far have been discovered via Doppler monitoring. Surprisingly, the fraction of multi-planetary systems discovered with each method is about the same, around 23\% according to the NASA Exoplanet Archive\footnote{ \url{https://exoplanetarchive.ipac.caltech.edu/}}, although the transit method can only discover those systems which are rather well aligned, while there is no strong bias against planets orbiting in tilted planes with respect to each other with the Doppler method. However, the apparent excess of single transiting systems has led some authors to speculate about the existence of a population of intrinsic singles or highly inclined multi-planet systems \citep{Lissauer2011ApJS..197....8L,Ballard2016ApJ...816...66B}.

These numbers are lower limits on the number of multi-planetary systems, because many planets presumably remain hidden even in the known systems because they are harder to detect due to smaller masses and/or larger periods. Furthermore, it has been shown that sparse radial velocity (RV) sampling especially of systems in 2:1 mean motion resonance (MMR) is prone to missing the second planet, and instead makes the system appear as if it hosts a single eccentric planet only \citep{anglada2010,kuerster2015,Boisvert2018MNRAS.480.2846B,Wittenmyer2019II}.

We carried out a Doppler survey for planets around 373 intermediate-mass evolved stars at Lick Observatory from 1999 to 2011 \citep{frink2001,2002ApJ...576..478F}, and are currently following up some of the most compelling systems with SONG \citep{2007CoAst.150..300G,2017ApJ...836..142G}. Here we report on one particular system from the Lick survey that was followed up with HARPS and SONG, the K1~III giant 7~CMa. 

One giant planet, namely 7~CMa~b, was reported to orbit 7~CMa already by \citet{Schwab2010}. It was independently found by \citet{2011ApJ...743..184W} based on RV data covering about one orbital cycle. Our data, covering about nine orbital cycles, indicate the presence of another giant planet in the system in a 4:3 MMR with the inner companion. Thus, 7~CMa is part of an elusive list of multi-planetary systems with giant host stars, some of which close to first-order MMRs, as discussed in \cite{Trifonov2019AJ....157...93T}. From the Lick sample, other MMR systems include $\eta$~Cet \citep[2:1,][]{trifonov2014} and $\nu$~Oph \citep[6:1,][]{quirrenbach2019}. 

Multi-planetary systems, and especially those in MMR, tell us much more about planet formation than single planet systems. Especially the formation of a 4:3 MMR is hard to explain \citep{rein2012} with current models, since the systems have to move through the 2:1 and 3:2 commensurabilities on their way to the 4:3 resonance, where the two planets are rather close together. 7~CMa is only the third system found in 4:3 MMR via Doppler monitoring, next to HD~200964 \citep{2011AJ....141...16J} and HD~5319 \citep{giguere2015}, involving massive, Jovian-like planets. All three systems are found around evolved host stars, in the late subgiant or early giant star phases, more massive than the Sun and with stellar radii in a narrow range between 4 and $5\,R_{\odot}$. Thus, more systems in the 4:3 MMR will certainly help to shed light on the formation mechanism of those and potentially other systems. 

The paper is organized as follows: Sect.~\ref{sec:star} is dedicated to the stellar parameters of 7~CMa, while in Sect.~\ref{sec:observ} we describe our RV data set. Section~\ref{sec:analysis} provides a dynamical analysis of the system, and in Sect.~\ref{sec:conclusions} we discuss the system and possible implications for its formation theory.


\section{Host star} \label{sec:star}

7~CMa ($=$~HD~47205, HIP~31592) is a bright K1 giant in the constellation Canis Major, accessible from most sites in both hemispheres. Comparing spectroscopic, photometric and astrometric observables to grids of stellar evolutionary models using Bayesian inference, \citet{stock18} derive an effective temperature of $T_{\mathrm{eff}} = 4826.0_{-55}^{+45}\,\mathrm{K}$ and surface gravity of $\log g = 3.19_{-0.07}^{+0.06}$. The metallicity was fixed to the value of $\mathrm{[Fe/H]}= 0.21 \pm 0.1$ from \cite{2007A&A...475.1003H}. The derived mass and radius of 7~CMa are $M = 1.34_{-0.12}^{+0.11}\,\mathrm{M_{\odot}}$ and $R = 4.87_{-0.14}^{+0.17}\,\mathrm{R_{\odot}}$. Table~\ref{tab:star} summarizes the main parameters of this star and previous values reported in the literature.

\begin{figure}
\centering
\includegraphics[width=\hsize]{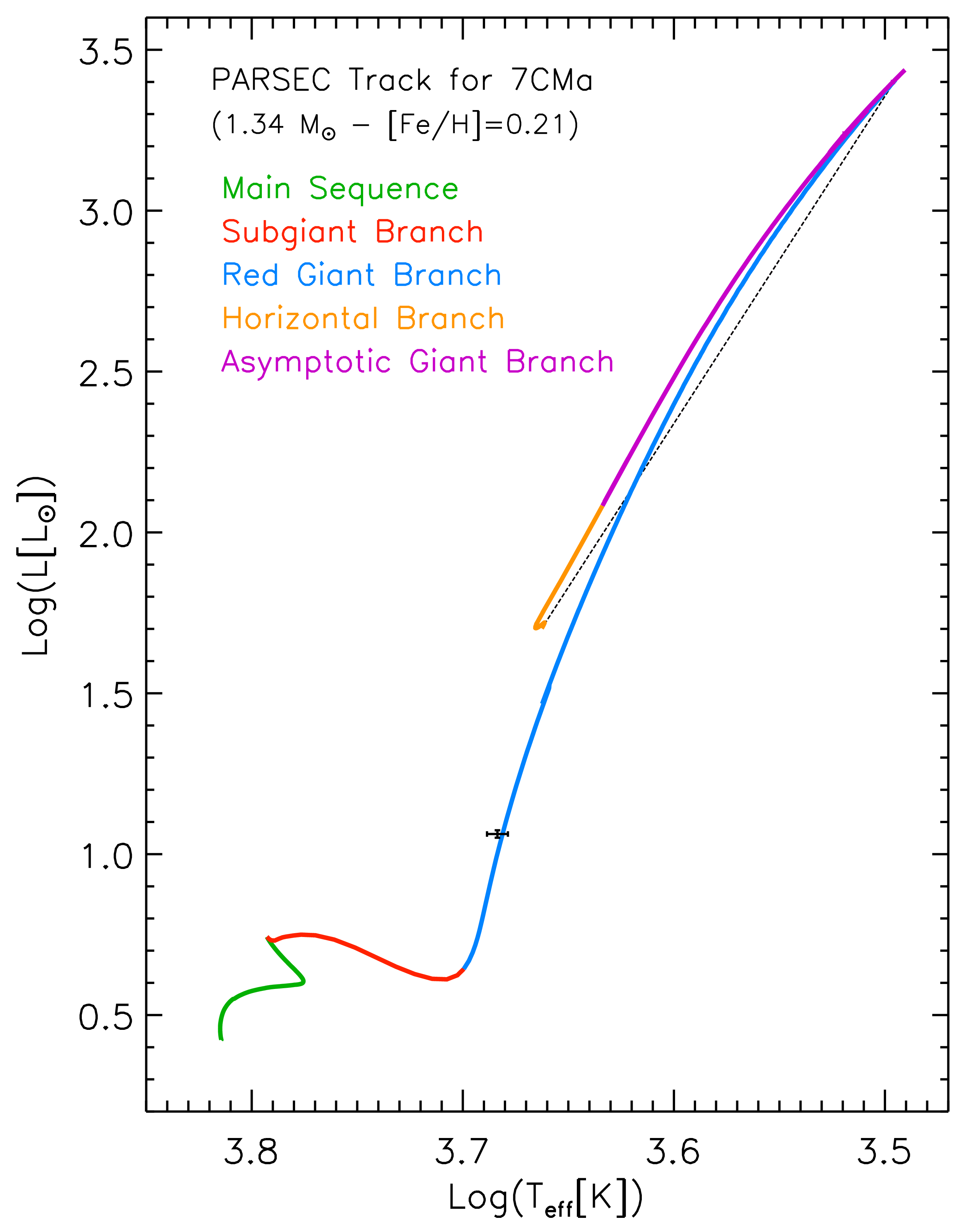}
\caption{Interpolated evolutionary track for 7 CMa in the Hertzsprung–Russell diagram. The different evolutionary phases are color-coded. The star's luminosity and temperature including uncertainties are shown in black.} \label{fig:evol}
\end{figure}

To illustrate the host star's expected evolutionary stage, we interpolate within the PARSEC grid of evolutionary tracks \citep{PARSEC} to obtain a track corresponding to the determined mass and metallicity of the star, which is shown in Fig.~\ref{fig:evol}. The momentary position of 7~CMa according to its temperature and luminosity is on the early ascent of the red giant branch (RGB), hence fusing hydrogen in a shell around an inert helium core. The star will undergo a helium flash, which happens on a time scale too short to be covered in the entries of the track. Therefore, the evolution along the black line running from the RGB tip to the beginning of the horizontal branch takes place quasi instantly.

\begin{table}
\centering
{\renewcommand{\arraystretch}{1.2}
 \footnotesize
\caption{Stellar parameters of 7~CMa.} \label{tab:star}
\begin{tabular}{lcr}
\hline\hline
\noalign{\smallskip}
Parameter 			    	& Value     		& Reference \\ 
\noalign{\smallskip}
\hline
\noalign{\smallskip}
Name                        & 7 CMa                     &      \\
HD                          & 47205                     &      \\
HIP                         & 31592                     & {\citet{2007A&A...474..653V}}     \\
$\alpha$                    & 06:36:41.03               & {\it Gaia} DR2     \\
$\delta$                    & --19:15:21.1              & {\it Gaia} DR2     \\
Spectral type				& K1\,III	    	        & {\citet{2006AJ....132..161G}}		\\
$V$ [mag] 				    & $3.91$	                & \citet{2002yCat.2237....0D}		\\
$d$ [pc]	    	    	& $19.81_{-0.16}^{+0.16}$	        & \citet{2018AJ....156...58B}		\\
\noalign{\smallskip}
\multicolumn{3}{c}{Photospheric parameters}\\
\noalign{\smallskip}
$T_{\mathrm{eff}}$ [K]		& $4826_{-55}^{+45}$	    & {\citet{stock18}}		\\
                    		& $4735_{-93}^{+35}$	    & {\it Gaia} DR2		\\
					        & $4792 \pm 100$	        & {\citet{2011ApJ...743..184W}}		\\
$\log g$				    & $3.19_{-0.07}^{+0.06}$	& {\citet{stock18}}		\\
					        & $3.25 \pm 0.10$	        & {\citet{2011ApJ...743..184W}}		\\
{[Fe/H]}				    & $0.21 \pm 0.1$	        & {\citet{2007A&A...475.1003H}}		\\
$v \sin i$ [$\mathrm{km\,s^{-1}}$]	& $1.15$	        & {\citet{2007A&A...475.1003H}}		\\            
\noalign{\smallskip}
\multicolumn{3}{c}{Derived physical parameters}\\
\noalign{\smallskip}
$M$ [M$_{\odot}$]			& $1.34_{-0.12}^{+0.11}$	& {\citet{stock18}}		\\
				    	    & $1.52 \pm 0.30$	        & {\citet{2011ApJ...743..184W}}		\\
$R$ [R$_{\odot}$]		    & $4.87_{-0.14}^{+0.17}$  	& {\citet{stock18}}		\\
        				    & $5.32_{-0.08}^{+0.21}$	& {\it Gaia} DR2		\\
					        & $2.3 \pm 0.1$		        & {\citet{2011ApJ...743..184W}}		\\
$L$ [L$_{\odot}$]	        & $11.55_{-0.20}^{+0.31}$	& {\citet{stock18}}		\\
                	        & $12.81_{-0.12}^{+0.12}$	& {\it Gaia} DR2		\\
Age [Gyr]				    & $4.3_{-1.3}^{+0.9}$       & {\citet{stock18}}		\\
\noalign{\smallskip}
\hline
\end{tabular}}
\tablebib{
    {\it Gaia} DR2: \citet{GaiaDR2}.
}
\end{table}

To date, one confirmed planet is already known to orbit 7~CMa. This planetary system was studied by \citet{Schwab2010} using Lick RVs and is the first reported discovery from the Pan-Pacific Planet Search survey \citep{2011ApJ...743..184W} at the 3.9\,m Anglo-Australian Telescope using the UCLES spectrograph. \citet{2011ApJ...743..184W} announced a giant planet ($m_b\sin i = 2.6\,\mathrm{M_{J}}$) with a period of $P_b = 763 \pm 17\,\mathrm{d}$ and eccentricity $e_b = 0.14 \pm 0.06$, based on 21 RV measurements taken between 2009 and 2011 with UCLES, adopting a stellar mass of $1.52 \pm 0.30\,\mathrm{M_{\odot}}$. Later, in a paper published together with another five discoveries, \citet{2016MNRAS.455.1398W} presented updated velocities and a refined orbit for 7~CMa together with six more measurements. The amplitude
of the Doppler signal is also confirmed to be independent of wavelength,
as expected for a planetary companion, by \citet{2015A&A...582A..54T} using near-infrared radial velocities obtained with CRIRES.


\section{Radial velocity measurements} \label{sec:observ}

We have collected RVs of 7~CMa from four different instruments during the last 19\,years as part of the project "Precise radial velocities of giant stars". In the following subsections, a short description of each instrument dataset is presented. The individual RVs from every instrument are listed in Table~\ref{tab:RVs}.

\subsection{Lick dataset} \label{subsec:lick}

Starting in 1999, our group carried out a radial velocity survey of 373 G- and K-giants at UCO/Lick Observatory using the 0.6\,m Coud\'e Auxiliary Telescope (CAT) together with the Hamilton Echelle Spectrograph with a nominal resolution of $R \sim 60\,000$ \citep[see, e.g.,][for a description of the survey and earlier results]{2002ApJ...576..478F,2006ApJ...652..661R}. Using the iodine cell method as described by \citet{1996PASP..108..500B} we obtained a typical RV precision of $\sigma_{\mathrm{jitt,Lick}} = 5$--$8\,\mathrm{m\,s^{-1}}$, adequate enough for our survey \citep{2015A&A...574A.116R}.

A total of 65 spectra for 7~CMa were taken between September 2000 and November 2011. The resulting RV measurements have a median precision of $\sim 5\,\mathrm{m\,s^{-1}}$.

\subsection{HARPS dataset} \label{subsec:harps}

We observed 7~CMa with the echelle optical spectrograph HARPS installed at the ESO 3.6\,m telescope at La Silla Observatory in Chile. We retrieved 11 measurements from 2006 to 2009 from the ESO Archive. Then, we triggered a campaign of 12 observations spanning three months in 2013, and 5 additional nights in March, April and September 2018. In this last campaign we took a series of consecutive exposures to study the intrinsic stellar variability (jitter) of the star. In total, 127 spectra were obtained. 

Radial velocities were obtained with the SERVAL program \citep{SERVAL} using high signal-to-noise templates created by co-adding all available spectra of the star. We split the HARPS data into two separate temporal subsets, HARPS-pre and HARPS-post, due to the HARPS fiber upgrade in June 2015, which introduced an RV offset that has to be modeled in the fitting process \citep{2015Msngr.162....9L}. In summary, a total of 25 nightly-averaged RVs (20 HARPS-pre and 5 HARPS-post) with a mean internal velocity uncertainty of $\sigma_{\mathrm{HARPS}} \sim 1\,\mathrm{m\,s^{-1}}$ were used in the analysis.

\subsection{SONG dataset} \label{subsec:song}

SONG (Stellar Observations Network Group) is planned as a network of 1\,m telescopes in both hemispheres that will carry out high-precision radial-velocity measurements of stars. The first node at Observatorio del Teide on Tenerife has been operating since 2014 and consists of the Hertzsprung SONG Telescope \citep{2014RMxAC..45...83A}. A total of 65 measurements were collected from 2015 to 2019 with a coud\'e echelle spectrograph through an iodine cell for precise wavelength calibration and radial-velocity determination \citep{2007CoAst.150..300G}. The data reduction pipeline is based on the IDL-routines of \citet{2002A&A...385.1095P} and the C{}\verb!++! reimplementation by \citet{2014PASP..126..170R}. More information about the data handling and RV extracting by the SONG collaboration can be found in \citet{2017ApJ...836..142G}. The typical uncertainties of the measurements are $\sigma_{\mathrm{SONG}} \sim 3\,\mathrm{m\,s^{-1}}$.

\subsection{UCLES dataset} \label{subsec:UCLES}

We also included the RVs obtained between 2009 and 2011 and published by \citet{2011ApJ...743..184W} together with six more measurements presented in \citet{2016MNRAS.455.1398W}. 

\section{Analysis} \label{sec:analysis}

\subsection{Periodogram search} \label{subsec:gls}

\begin{figure}
\centering
\includegraphics[width=\hsize]{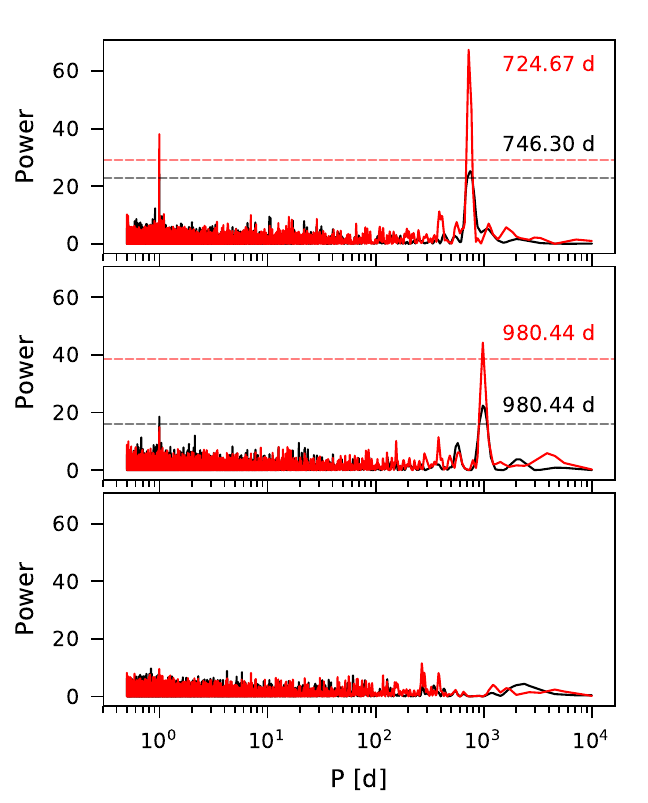}
\caption{
 \textit{Top panel}: Lomb-Scargle (LS) periodogram of the RVs. The periodogram of the Lick data only is shown in black, while the LS periodogram of the complete RV dataset is shown in red. A radial velocity offset and a jitter term are individually fitted for each dataset in adittion to a global linear trend. The highest peak at $\sim 735\pm10$\,d is consistent with the planet claimed by \citet{2011ApJ...743..184W}. \textit{Center panel}: LS periodogram of the residuals to the Keplerian orbit fit of the $\sim 735$\,d signal. The highest peak at 980\,d hints at the presence of a second planet in the system. \textit{Bottom panel}: LS periodogram of the residuals to the Keplerian orbital fit of the two main signals. The horizontal lines in the panels show a false alarm probability level of 0.1\% in black and red for Lick and complete dataset, respectively. The period of the highest peak in the LS periodogram is indicated with the color corresponding to each of the datasets.} \label{fig:period_lick}
\end{figure}

We compute a Lomb-Scargle \citep[LS;][]{Lomb76,Scargle82} periodogram to look for periodic signals in the RV data. Using the Lick data alone, we find a highly significant peak around 746\,d, as shown in black in the top panel of Fig.~\ref{fig:period_lick}. This result is consistent with the already known planet of the system, where \citet{2011ApJ...743..184W} announced a signal of $763\,\mathrm{d}$. However, after fitting the reported planetary signal in our data, a significant peak around $\sim 980\,\mathrm{d}$ is found in the residuals with false alarm probability smaller than 0.1\% (see central panel of Fig.~\ref{fig:period_lick}). After the fitting of these two periodic signals, no further signals are evident in the residuals, as shown in the bottom panel of Fig.~\ref{fig:period_lick}. The weighted root-mean-squared of the residuals improves from $12.9\,\mathrm{m\,s^{-1}}$ in the one-planet model fit to $8.2\,\mathrm{m\,s^{-1}}$ in the two-planet. The second signal could only be revealed thanks to the longer timespan of the Lick RVs compared to UCLES.

Following the second planet hint in the Lick data, we collected more observations with different facilities. The LS periodogram of the complete RV data set shows narrower and stronger signals at the aforementioned periods, as shown in red in Fig.~\ref{fig:period_lick}, further supporting the second planet hypothesis and constraining its orbital properties. The period of the second planet at $980$\,d is nearly in a 4:3 ratio with the first companion, suggesting a two-planet system likely in orbital resonance. 

\subsection{Keplerian and dynamical modeling} \label{subsec:bestfit}

\begin{figure*}
\centering
\includegraphics[width=\hsize]{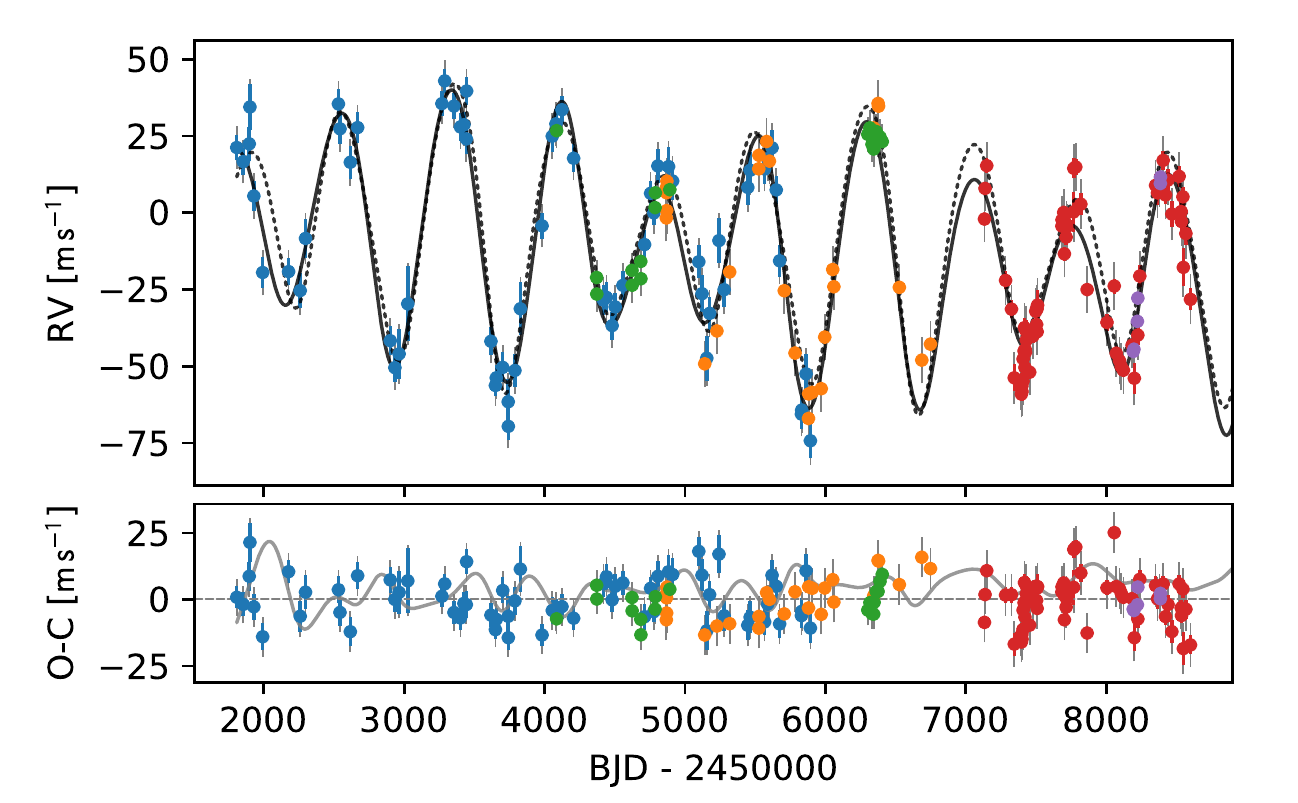}
\caption{
Time series of the 182 RVs obtained for 7~CMa from September 2000 to April 2019 with Lick (blue), UCLES (orange), HARPS (before/after the fibre upgrade of 2015 in green/purple, respectively), and SONG (red) facilities. The vertical gray lines mark the error bars including jitter. The best double Keplerian fit to the data is drawn with a dotted line, while the solid black line indicates the best dynamical two-planet fit. The residuals of the dynamical fit and the difference between the Keplerian and dynamical models (solid gray line) are shown in the bottom panel.} \label{fig:dyn_fit}
\end{figure*}

We adopt a maximum-likelihood estimator coupled with a downhill simplex algorithm \citep{NelderMead65,1992nrfa.book.....P} to determine the orbital parameters of the planet candidates orbiting 7~CMa. The negative logarithm of the model's likelihood function ($-\ln\mathcal{L}$) is minimized while optimizing the planet orbital parameters (RV semi-amplitudes $K_{b,c}$, periods $P_{b,c}$, eccentricities $e_{b,c}$, arguments of periastron $\omega_{b,c}$, mean anomalies $M_{b,c}$, a RV zero-point offset for each dataset and a global RV slope). We also include the RV instrumental jitter as an additional model parameter for each dataset. Afterward, we estimate the uncertainties of the best-fit parameters using the Markov Chain Monte Carlo (MCMC) sampler \texttt{emcee} \citep{emcee}. We adopt flat priors for all parameters and select the 68.3 confidence interval levels of the posterior distributions as $1\sigma$ uncertainties. We use the {\sc Exo-Striker} \citep{exostriker} to perform all analyses discussed here.

First, we fit the RV dataset with a double-Keplerian model. The relatively close planetary orbits and the derived minimum masses of the planets indicate that the planets will have relatively close encounters during the time of the observations, which may be detected in our data. Therefore, a more appropriate N-body dynamical model is applied, which takes into account the gravitational interactions between the massive bodies by integrating the equations of motion using the Gragg-Bulirsch-Stoer method \citep{1992nrfa.book.....P}. For consistency with the unperturbed Keplerian frame and in order to work with minimum dynamical masses, we assume an edge-on and coplanar configuration for the 7~CMa system (i.e. $i_{b,c} = 90\,\mathrm{deg}$ and $\Delta\Omega = 0\,\mathrm{deg}$). The timestep employed in the integration is 1\,d.

Figure~\ref{fig:dyn_fit} shows the best-fit solutions from each of the schemes together with the complete RV dataset. The 7~CMa system contains two Jupiter-like planets with minimum masses $m_b \sin i \sim 1.8\,M_J$ and $m_c \sin i \sim 0.9\,M_J$ orbiting in low-eccentricity orbits. The period ratio of the planets in the 7~CMa system is close to 1.33, potentially trapped in a 4:3 mean motion resonance. The two models are almost equivalent and taking the gravitational interactions into account in the fitting does not turn into a significant improvement in the $-\ln\mathcal{L}$ of the fit with respect to the Keplerian model. The relatively short span of the observations ($\sim 9$ orbits) is not enough to detect the secular perturbation of the orbits. However, although a double-Keplerian or a full self-consistent N-body dynamical model fit the RV data almost equally well, we decide to base our analysis on the dynamical model which, given the derived close orbits and the Jovian-like masses of the planets, is better justified. The orbital parameters of the two planets for the dynamical best-fit and the posterior distributions from the MCMC sampling are summarized in Table~\ref{tab:best_fit_dyn}.

We note that the jitter values derived for each dataset are all above $5\,\mathrm{m\,s^{-1}}$, as expected for K giants due to p-mode oscillations  \citep{2006A&A...454..943H,2008A&A...480..215H}. 
Studying our high-cadence HARPS data we measure a peak-to-peak variation in the RVs of $5\,\mathrm{m\,s^{-1}}$ on timescales of the order of 30\,min. From the scaling relations of \citet{Kjeldsen2011A&A...529L...8K} we expect a velocity jitter of between 3--$5\,\mathrm{m\,s^{-1}}$ for 7~CMa due to $p$-mode oscillations, fully consistent with the derived jitter terms.

\begin{table*}
    
\centering   
\caption{Orbital parameters of the 7~CMa system. The first column shows the mode of the dynamical MCMC samples and the 68\% credibility intervals as errorbars. The medium and last columns show the nominal and stable best dynamical fits, respectively. All fits are fixed to be edge-on and coplanar ($i_{b,c} = 90\,\mathrm{deg}$, $\Omega_b = \Omega_c = 0\,\mathrm{deg}$). We use the JD of the first RV observation, $\mathrm{JD}=2451808.021$, to set the epoch.} \label{tab:best_fit_dyn}

\begin{tabular}{l@{\hskip 0.3in}rr@{\hskip 0.3in}rr@{\hskip 0.3in}rr}     
    
    \hline\hline
    \noalign{\smallskip}     
    ~~~~~~~~~                         & \multicolumn{2}{c}{MCMC samples} & \multicolumn{2}{c}{Best-fit} & \multicolumn{2}{c}{Stable best-fit} \\
    \noalign{\smallskip}     
    Parameter                         & 7 CMa b & 7 CMa c & 7 CMa b & 7 CMa c & 7 CMa b & 7 CMa c \\
    \hline
    \noalign{\smallskip}
        
        $K$  [m\,s$^{-1}$]            &     34.3$_{-0.9}^{+1.2}$       &     14.9$_{-1.1}^{+0.9}$ &     32.9       &     14.8 &     35.1       &     15.2 \\ \noalign{\vskip 0.9mm}
        $P$  [d]                      &    735.1$_{-1.0}^{+14.8}$      &    996.0$_{-52.4}^{+1.5}$ &    758.5      &    925.5 &    736.9      &    988.9 \\ \noalign{\vskip 0.9mm}
        $e$                           &    0.06$_{-0.03}^{+0.03}$    &    0.08$_{-0.04}^{+0.05}$ &    0.055    &    0.075 &   0.055    &    0.046 \\ \noalign{\vskip 0.9mm}
        $\omega$  [deg]               &    165.3$_{-70.8}^{+5.1}$      &    233.5$_{-40.2}^{+7.7}$ &    111.9      &    240.7 &    116.4      &    226.6 \\ \noalign{\vskip 0.9mm}
        $M_{\rm 0}$  [deg]            &    154.6$_{-0.8}^{+85.8}$       &    306.0$_{-21.1}^{+19.1}$ &    237.8       &    260.8 &    216.5       &    308.8 \\ \noalign{\vskip 0.9mm}
    \noalign{\smallskip}
        $a$  [au]                     &      1.758$_{-0.001}^{+0.024}$      &      2.153$_{-0.08}^{-0.003}$ &      1.795      &      2.050 &      1.761      &      2.143 \\ \noalign{\vskip 0.9mm} 
        $m \sin i$  [$M_J$]           &      1.85$_{-0.04}^{+0.06}$      &      0.87$_{-0.06}^{+0.06}$ &      1.798      &      0.862 &      1.895      &      0.906 \\ \noalign{\vskip 0.9mm} 
    \noalign{\smallskip}
        $\dot{\gamma}$ [$\mathrm{m\,s^{-1}\,d^{-1}}$]                                               &     \multicolumn{2}{c}{-0.0025$_{-0.0006}^{+0.0006}$} &     \multicolumn{2}{c}{-0.0036} &     \multicolumn{2}{c}{-0.0032} \\ \noalign{\vskip 0.9mm}
        $\gamma_{\mathrm{Lick}}$ [$\mathrm{m\,s^{-1}}$]             &     \multicolumn{2}{c}{8.7$_{-2.1}^{+2.1}$} &     \multicolumn{2}{c}{11.1} &     \multicolumn{2}{c}{9.2} \\ \noalign{\vskip 0.9mm}
        $\gamma_{\mathrm{UCLES}}$ [$\mathrm{m\,s^{-1}}$]            &     \multicolumn{2}{c}{16.8$_{-3.5}^{+3.5}$} &     \multicolumn{2}{c}{20.1} &     \multicolumn{2}{c}{19.1} \\ \noalign{\vskip 0.9mm}
        $\gamma_{\mathrm{HARPS-pre}}$ [$\mathrm{m\,s^{-1}}$]        &     \multicolumn{2}{c}{17.0$_{-2.9}^{+2.9}$} &     \multicolumn{2}{c}{21.5} &     \multicolumn{2}{c}{18.8} \\ \noalign{\vskip 0.9mm}
        $\gamma_{\mathrm{HARPS-post}}$ [$\mathrm{m\,s^{-1}}$]       &     \multicolumn{2}{c}{35.3$_{-5.4}^{+5.7}$} &     \multicolumn{2}{c}{42.9} &     \multicolumn{2}{c}{37.6} \\ \noalign{\vskip 0.9mm}
        $\gamma_{\mathrm{SONG}}$ [$\mathrm{m\,s^{-1}}$]             &     \multicolumn{2}{c}{2737.5$_{-4.6}^{+4.3}$} &     \multicolumn{2}{c}{2744.9} &     \multicolumn{2}{c}{2741.0} \\ \noalign{\vskip 0.9mm}
        $\sigma_{\mathrm{jitt,Lick}}$ [$\mathrm{m\,s^{-1}}$]        &     \multicolumn{2}{c}{7.6$_{-1.3}^{+1.4}$} &     \multicolumn{2}{c}{5.5} &     \multicolumn{2}{c}{6.8} \\ \noalign{\vskip 0.9mm}
        $\sigma_{\mathrm{jitt,UCLES}}$ [$\mathrm{m\,s^{-1}}$]       &     \multicolumn{2}{c}{8.1$_{-1.2}^{+1.4}$} &     \multicolumn{2}{c}{7.5} &     \multicolumn{2}{c}{8.0} \\ \noalign{\vskip 0.9mm}
        $\sigma_{\mathrm{jitt,HARPS-pre}}$ [$\mathrm{m\,s^{-1}}$]   &     \multicolumn{2}{c}{5.9$_{-1.1}^{+1.6}$} &     \multicolumn{2}{c}{5.7} &     \multicolumn{2}{c}{5.8} \\ \noalign{\vskip 0.9mm}
        $\sigma_{\mathrm{jitt,HARPS-post}}$ [$\mathrm{m\,s^{-1}}$]  &     \multicolumn{2}{c}{5.8$_{-2.0}^{+3.6}$} &     \multicolumn{2}{c}{4.9} &     \multicolumn{2}{c}{3.6} \\ \noalign{\vskip 0.9mm}
        $\sigma_{\mathrm{jitt,SONG}}$ [$\mathrm{m\,s^{-1}}$]        &     \multicolumn{2}{c}{8.7$_{-1.0}^{+1.2}$} &     \multicolumn{2}{c}{7.3} &     \multicolumn{2}{c}{8.6} \\ \noalign{\vskip 0.9mm}
    \noalign{\smallskip}
        $-\ln\mathcal{L}$             &   \multicolumn{2}{c}{\dots} &   \multicolumn{2}{c}{-631.679} &   \multicolumn{2}{c}{-637.705} \\
        N$_{\rm RV}$ data             &        \multicolumn{2}{c}{182} &        \multicolumn{2}{c}{182} &        \multicolumn{2}{c}{182} \\
    \noalign{\smallskip}
    \hline

\end{tabular}  
    
    
    
\end{table*}

\subsection{System configuration} \label{subsec:mcmc_stab}

The MCMC analysis provides a median solution which is in agreement with the best-fit solution except for the periods of both planets. Moreover, both solutions fail to preserve stability in a short period of time, compatible with a handful of orbits of the outer planet. Thus, requiring long-term stability can further constrain the system configuration. The formally best-fit solution does not necessarily have to be stable, but we should find stable configurations close to the formally best-fit solution.

To test the stability of the planetary system around 7~CMa, we integrate each individual MCMC sample using the Wisdom-Holman symplectic algorithm (MVS) integrator contained in the SWIFT package \citep{1998AJ....116.2067D}. This is a symplectic algorithm created to perform long-term numerical orbital integrations of solar system objects. All samples have been integrated for 1\,Myr and the time step used for the integrations is 1\,d, to ensure accurate temporal resolution. A stable system is defined if none of the planets are ejected or experience a collision, the semi-major axes remain within 10\% from the initial values, and the eccentricities are lower than 0.95 - which otherwise would lead to nonphysical orbits inside the star's radius - during the complete integration time.

\begin{figure*}[ht!]
\centering
		\includegraphics[width=0.99\textwidth]{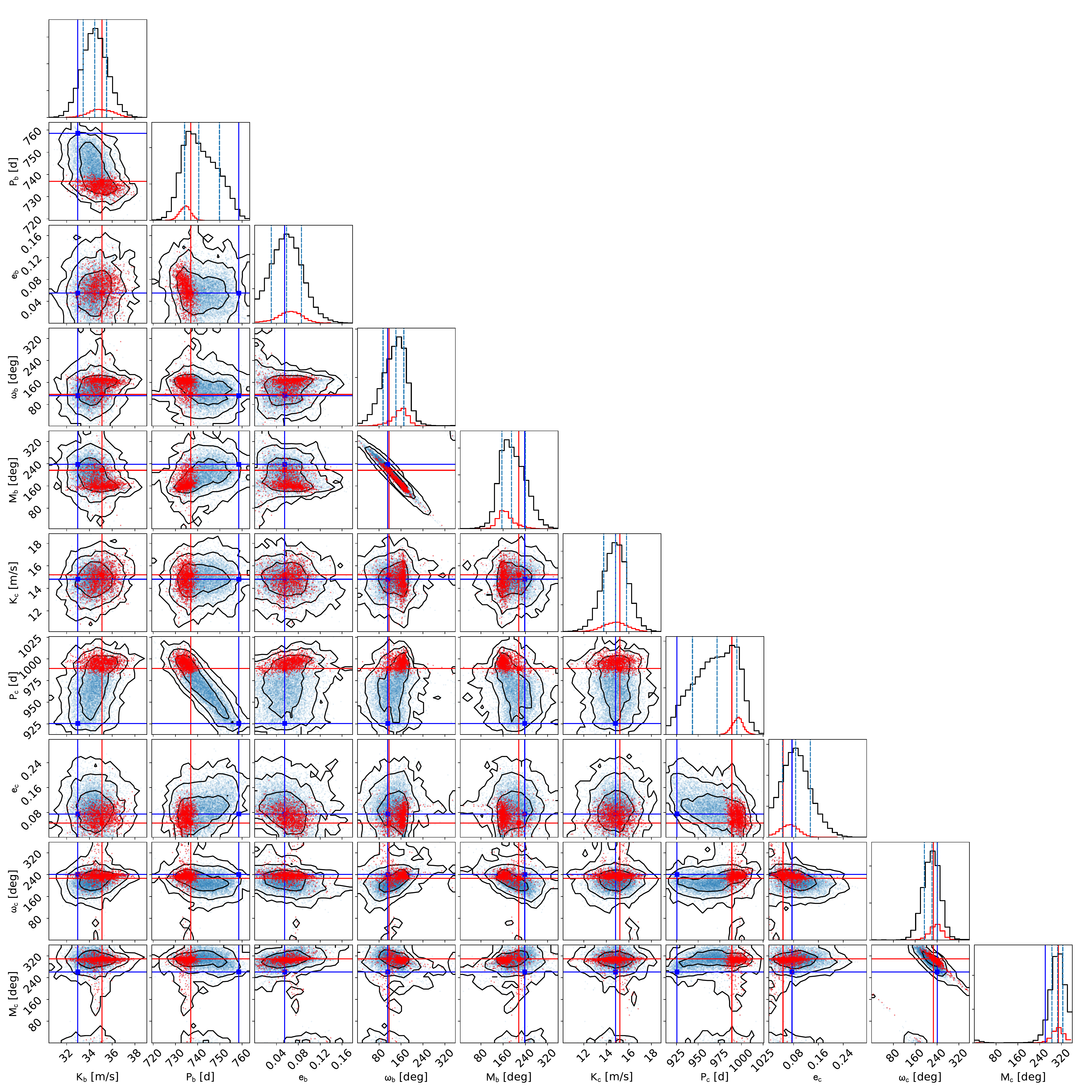}
		\caption{Posterior distributions of the orbital parameters of the 7~CMa system. Each panel contains $\sim 100\,000$ samples which are tested for 1\,Myr dynamical stability using the MVS integrator. Stable solutions are overplotted in red. The upper panels of the corner plot show the probability density distributions of each orbital parameter of the overall MCMC samples (black) and the stable ones (red). The vertical dashed lines mark the 16th, 50th and the 84th percentiles of the the overall MCMC samples. Contours are drawn to improve the visualization of the two-dimensional histograms and indicate the 68.3\%, 95.5\%, and 99.7\% confidence interval levels (i.e. 1$\sigma$, 2$\sigma$ and 3$\sigma$). Blue and red crosses indicate the dynamical best-fit solution (central column of Table~\ref{tab:best_fit_dyn}) and the stable best-fit solution (last column of Table~\ref{tab:best_fit_dyn}), respectively.} \label{fig:mcmc_stab}
\end{figure*}

Figure~\ref{fig:mcmc_stab} shows the posterior MCMC distribution of the orbital parameters using a dynamical, edge-on, coplanar model. The histogram panels on the top of Fig.~\ref{fig:mcmc_stab} provide a comparison between the probability density function of the complete MCMC samples (blue) and the samples  that are stable for at least 1\,Myr (red) for each fitted parameter. The corner-plot panels represent all possible parameter combinations with respect to the best dynamical fit from Table~\ref{tab:best_fit_dyn}, whose position is marked with a blue cross. The black 2D contours are constructed from the overall MCMC samples and indicate the 68.3\%, 95.5\%, and 99.7\% confidence interval levels (i.e. 1$\sigma$, 2$\sigma$ and 3$\sigma$). For clarity, in Fig. 6 the stable samples are overplotted in red and the stable solution with the maximum $\ln \mathcal{L}$ is marked with a red cross.

We find that $\sim 13\%$ of the MCMC samples are stable. Moreover, the mode of the overall and stable samples are coincident for every orbital parameter and the best-fit stable solution is almost coincident with the median of the posteriors. Therefore, although the nominal best-fit solution ($\ln \mathcal{L} = -631.7$) derives a period for planets b and c that are off by $2\sigma$ from the mode of the samples, the actual configuration of the 7~CMa system is better represented by the stable best-fit solution ($\ln \mathcal{L} = -637.7$) shown in the last column of Table~\ref{tab:best_fit_dyn}. 

In this stable configuration, the orbital periods of the planets are $P_b \sim 737$\,d and $P_c \sim 989$\,d, implying a period ratio of 1.34; while the nominal best-fit solution has a period ratio $P_c / P_b = 1.22$. This value is far from the 4:3 value that can preserve the stability of the system by trapping the planets in a 4:3 MMR, preventing the planets from close encounters. While the posterior distributions for the planet periods in the overall MCMC samples are very wide and asymmetrical, the periods of the stable samples are narrow and Gaussian-like, further supporting the validity of this solution despite its slightly lower statistical significance.

Furthermore, to describe correctly the data it is necessary to include a linear trend in the RV models. The RV slope is particularly evident in the Lick dataset and corresponds to $\sim 1\,\mathrm{m\,s^{-1}}$ per year. A planet in a 50\,yr-period circular orbit assuming an RV semi-amplitude of about $12.5\,\mathrm{m\,s^{-1}}$ (which corresponds to the $1\,\mathrm{m\,s^{-1}\,yr^{-1}}$ trend over 25\,yr) would have a minimum mass of $2\,M_J$. Increasing the period and semi-amplitude by a factor 2 the minimum mass would be $20\,M_J$. On the other hand, a three-planet fit to the data yields a $\ln \mathcal{L}$ indistinguishable from a two-planet model and the period of the candidate is not well constrained. Although a third planet would not affect the stability of the inner pair, it could play an important role in the formation history of the system. Long-cadence observations of 7~CMa with the same instrumentation will shed light into the nature of the linear trend and possible further companions in the system.

Last, we tested the impact of coplanar inclined orbits (i.e., $i_{b,c} \neq 90\,\mathrm{deg}$, $\Omega_b = \Omega_c = 0\,\mathrm{deg}$) in the stability of the system. The impact of the inclinations with respect to the observer's line of sight mainly manifests itself through the derived planetary masses, which are increased by a factor $\sin i$. We chose a random subset of stable samples covering the parameter space of the red points in Fig.~\ref{fig:mcmc_stab} and integrate them for 10\,Myr with inclinations varying from 0 to 90\,deg. We choose this simpler approach since a complete sampling of coplanar and mutually inclined systems (with $i_b$, $i_c$, $\Omega_b$, and $\Omega_c$ as free parameters) in a dynamical fashion is computationally very expensive. Our analysis show that these stable solutions cannot even preserve stability on very short timescales for $i_b=i_c \lesssim 70\,\mathrm{deg}$. The larger planetary masses and higher interaction rate make these solutions much more fragile than the edge-on coplanar system.

\subsection{Dynamical properties} \label{subsec:stab}

A period ratio close to 1.33 does not ensure that the system is indeed trapped in a 4:3 MMR. To test this scenario it is necessary to study the long-term evolution of the orbital parameters and, particularly, the resonant angles. For the 4:3 MMR, these angles are defined as:
\begin{equation}
\begin{split}
\sigma_b &= 4 \lambda_c - 3 \lambda_b - \omega_b \\
\sigma_c &= 4 \lambda_c - 3 \lambda_b - \omega_c \quad ,
\end{split}
\end{equation}
where the mean longitude $\lambda_i = M_i + \omega_i$ \citep[see, e.g.,][]{MurrayDermott1999ssd..book.....M}.

\begin{figure*}
	\centering
	\includegraphics[width=0.48\hsize]{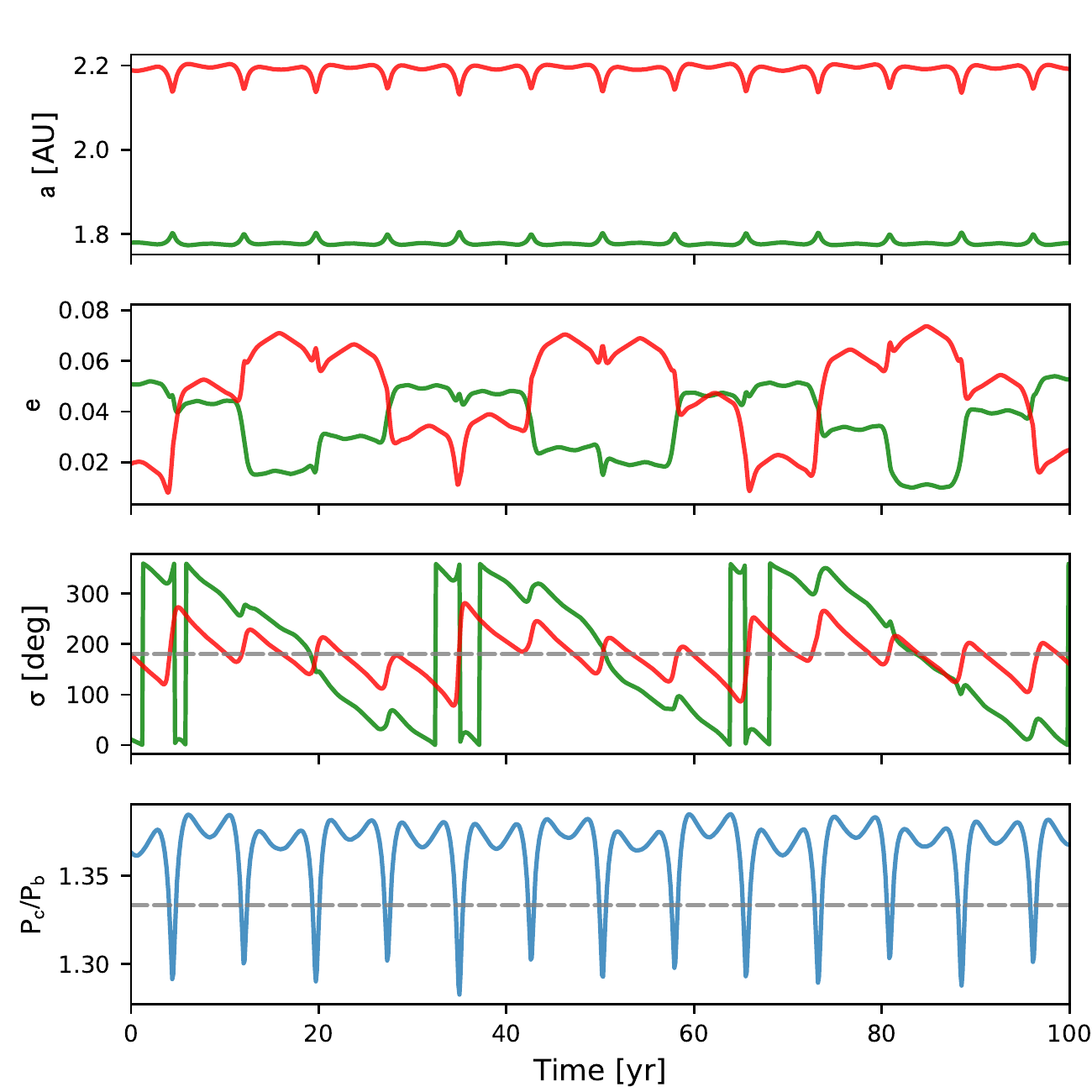}
	\includegraphics[width=0.48\hsize]{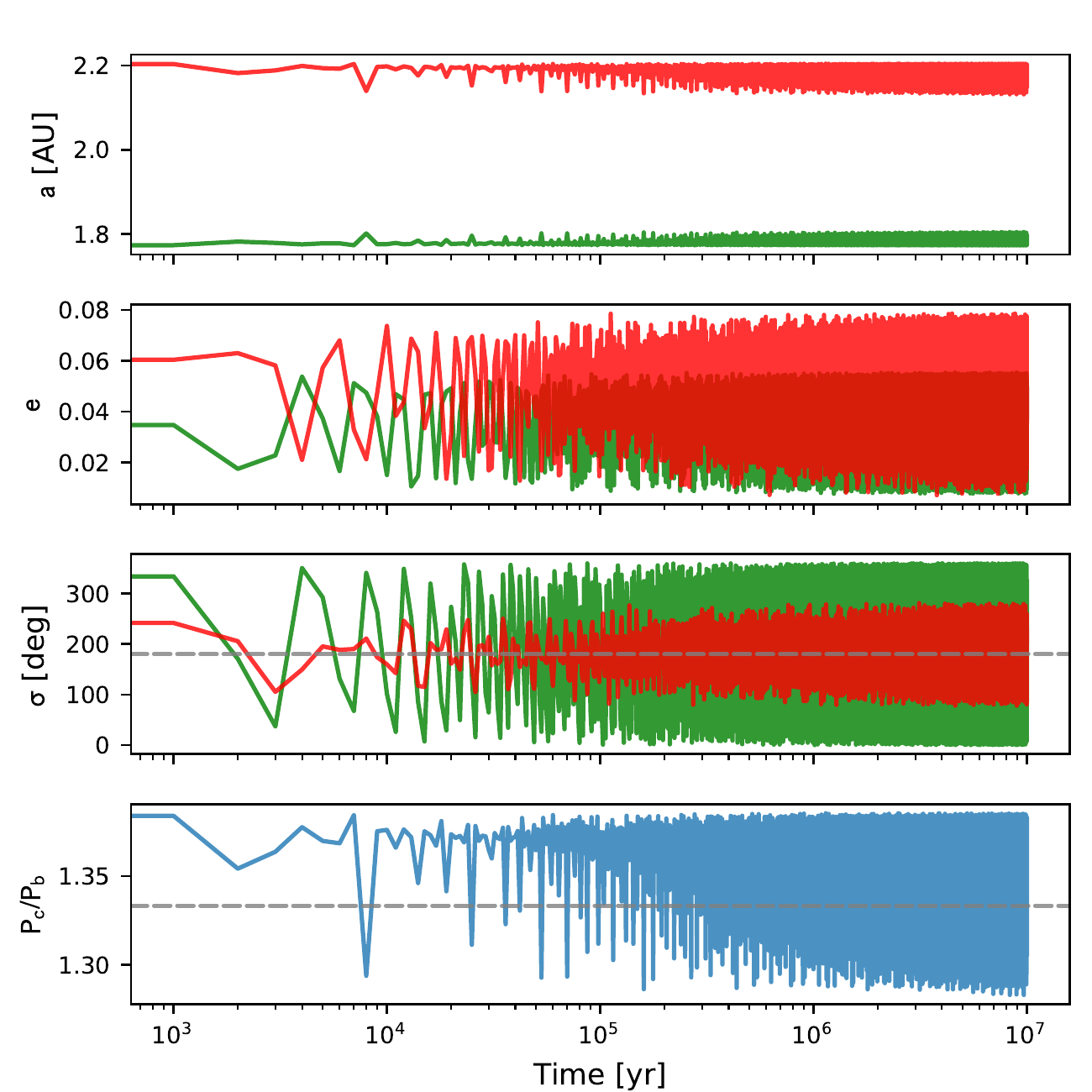}
	\caption{
		Semi-major axes, eccentricities, resonant angles and period-ratio evolution of one of the stable fits for 10\,Myr. Planet b is shown in green, while planet c is in red. Left panel shows a 100\,yr zoomed region of the complete 10\,Myr simulation, shown in the right panel (note the logarithmic scale in the X-axis). The resonant angle of the first planet is circulating, while the second is librating around $180\si{\degree}$, confirming that the system is trapped in a 4:3 MMR configuration. The system suffers from strong gravitational interactions on very short timescales, but it can preserve stability for 10\,Myr.
	} \label{fig:long_stab}
\end{figure*}

Figure~\ref{fig:long_stab} shows the long-term evolution of an arbitrary stable sample chosen from the MCMC. This solution can preserve stability for at least 10\,Myr, with semi-major axes and eccentricities oscillating rapidly with period ratios close to 1.33. The semi-major axes are strongly constrained to $a_b \sim 1.8\,\mathrm{au}$ and $a_c \sim 2.2\,\mathrm{au}$, while the period-ratio of the planets oscillates slightly above the 4:3 value, marked with a gray dashed-line. The periodic drops in $P_c/P_b$ are a consequence of the rapid variations in the semi-major axes when the two planets get close to each other.

The behavior of the resonant angles defines the location of the system with respect to the resonance: when one of the angles is librating, the system is said to be inside the resonance. The resonant angle of the first planet circulates from $0\si{\degree}$ to $360\si{\degree}$, while $\sigma_c$ is librating around $180\si{\degree}$. The confinement of  $\sigma_c$ around $180\si{\degree}$ shows that it is the truly resonant librating angle of the 7~CMa system, as shown previously for the HD~200964 system by \citet{2015A&A...573A..94T}. Therefore, we can conclude that the two-planet system is effectively trapped in the narrow stable region of the 4:3 mean motion resonance and that the stability analysis reveals the true configuration of the system.

\section{Conclusions} \label{sec:conclusions}

We report the discovery of a second planet orbiting the K-giant star 7~CMa. The extensive RV dataset reveals two massive Jupiter-like planets ($m_b \sin i \approx 1.9\,M_{J}$, $m_c \sin i \approx 0.9\,M_{J}$) orbiting closely in 4:3 MMR around their parent star. We find the true configuration of the system by studying the long-term stability of the planets since with the current data the periods are not well constrained. The mode of the MCMC samples is coincident with the median of the stable samples, which are narrow and Gaussian-like for all orbital parameters. The best nominal solution is within $2\sigma$ from the mode of the MCMC samples and with $\Delta \ln \mathcal{L} = 6$ with respect to the best stable fit. 

The two-planet system around 7~CMa is the third to be discovered in 4:3 MMR, after HD~200964 \citep{2011AJ....141...16J} and HD~5319 \citep{giguere2015}. The existence of these massive planet systems challenge formation models since migration scales for passing through the 2:1 and 3:2 resonances are extremely short. This migration speed is almost impossible to achieve because a large amount of angular momentum should be delivered to the disk, as pointed out by \citet{Ogihara2013ApJ...775...34O}. \citet{rein2012} reached the same conclusions using hydrodynamical simulations of convergent migration and in-situ formation. On the other hand, \citet{2015A&A...573A..94T} were able to reproduce the formation process of HD~200964 using models that contained an interaction between the type I and type II of migration, planetary growth and stellar evolution from the main sequence to the sub-giant branch. However, the authors pointed out that the formation process is very sensitive to the planetary masses and protoplanetary disk parameters, where only a thin, vertically isothermal and laminar disk, with a nearly constant surface density profile allows the embryo-sized planets to reach the 4:3 resonant configuration. Another possible escape from the 2:1 and 3:2 resonances could be resonance overstability, as proposed by \citet{Goldreich2014AJ....147...32G}. In this case, convergent planetary migration with strongly damped eccentricities may only lead to a temporal capture at the 2:1 and 3:2 resonances. A detailed analysis on the formation of the 7~CMa system is out of the scope of this paper, but, given the similarities in the mass-ratio of the planets and the host star we believe that this system could have undergone a similar formation and evolution as the HD~200964 system.


\begin{acknowledgements}
R.\,L. has received funding from the European Union’s Horizon 2020 research and innovation program under the Marie Skłodowska-Curie grant agreement No.~713673 and financial support through the “la Caixa” INPhINIT Fellowship Grant LCF/BQ/IN17/11620033 for Doctoral studies at Spanish Research Centres of Excellence from “la Caixa” Banking Foundation, Barcelona, Spain.

S.\,R.\ and V.\,W.\ acknowledge support of the DFG priority program SPP 1992 "Exploring the Diversity of Extrasolar Planets" (RE 2694/5-1). M.H.L. was supported in part by Hong Kong RGC grant HKU 17305618.

Based on observations made with the Hertzsprung SONG telescope operated on the Spanish Observatorio del Teide on the island of Tenerife by the Aarhus and Copenhagen Universities and by the Instituto de Astrof{\'i}sica de Canarias.

\end{acknowledgements}

\bibliographystyle{aa} 
\bibliography{biblio} 


\begin{appendix} 

\section{Radial velocity measurements}

\longtab[1]{
    \begin{longtable}{lrcr}
        \caption{\label{tab:RVs} Radial velocities and formal uncertainties of 7~CMa.}\\ 
        \hline
        \hline
        \noalign{\smallskip}
        BJD        & RV                     & $\sigma_{\rm RV}$ & Instrument\\
                   & ($\mathrm{m\,s^{-1}}$) & ($\mathrm{m\,s^{-1}}$) & \\
        \noalign{\smallskip}
        \hline
        \endfirsthead
        \caption{RVs (cont.)}\\
        \hline
        \hline
        \noalign{\smallskip}
        BJD        & RV                     & $\sigma_{\rm RV}$ & Instrument\\
                   & ($\mathrm{m\,s^{-1}}$) & ($\mathrm{m\,s^{-1}}$) & \\
        \noalign{\smallskip}
       \hline
        \noalign{\smallskip}
       \endhead
        \noalign{\smallskip}
       \hline
       \endfoot
    \noalign{\smallskip}  
2451808.021 & 32.3 & 4.2 & Lick \\
2451853.99 & 27.7 & 4.5 & Lick \\
2451896.855 & 33.5 & 5.5 & Lick \\
2451901.86 & 45.5 & 7.4 & Lick \\
2451929.723 & 16.5 & 5.0 & Lick \\
2451992.658 & -8.4 & 5.0 & Lick \\
2452177.026 & -8.1 & 4.4 & Lick \\
2452259.801 & -14.3 & 5.8 & Lick \\
2452297.795 & 2.7 & 6.3 & Lick \\
2452531.995 & 46.5 & 5.0 & Lick \\
2452543.041 & 38.4 & 5.0 & Lick \\
2452616.817 & 27.5 & 5.4 & Lick \\
2452668.714 & 38.8 & 5.1 & Lick \\
2452901.024 & -30.6 & 4.8 & Lick \\
2452933.909 & -39.4 & 4.7 & Lick \\
2452963.961 & -34.9 & 8.1 & Lick \\
2453025.82 & -18.6 & 12.2 & Lick \\
2453269.05 & 46.6 & 4.0 & Lick \\
2453288.997 & 54.0 & 3.8 & Lick \\
2453354.761 & 45.8 & 4.3 & Lick \\
2453400.787 & 39.0 & 4.6 & Lick \\
2453425.719 & 39.7 & 4.2 & Lick \\
2453442.655 & 35.1 & 5.0 & Lick \\
2453444.634 & 50.7 & 4.6 & Lick \\
2453618.029 & -30.8 & 4.6 & Lick \\
2453650.022 & -45.2 & 3.5 & Lick \\
2453656.062 & -42.7 & 4.3 & Lick \\
2453701.917 & -39.3 & 4.9 & Lick \\
2453740.972 & -50.5 & 7.5 & Lick \\
2453741.835 & -58.5 & 4.6 & Lick \\
2453788.657 & -40.3 & 5.3 & Lick \\
2453827.648 & -20.2 & 8.6 & Lick \\
2453982.026 & 6.8 & 4.3 & Lick \\
2454054.911 & 36.0 & 5.1 & Lick \\
2454080.991 & 40.0 & 4.7 & Lick \\
2454085.648 & 48.2 & 0.3 & HARPS-pre \\
2454123.821 & 44.6 & 4.6 & Lick \\
2454206.666 & 28.8 & 4.5 & Lick \\
2454370.862 & 0.4 & 0.3 & HARPS-pre \\
2454371.828 & -5.0 & 0.3 & HARPS-pre \\
2454418.951 & -17.4 & 5.0 & Lick \\
2454440.83 & -16.5 & 4.9 & Lick \\
2454480.882 & -25.7 & 4.6 & Lick \\
2454502.828 & -19.6 & 4.5 & Lick \\
2454557.664 & -12.6 & 4.6 & Lick \\
2454622.437 & 2.7 & 0.4 & HARPS-pre \\
2454623.454 & -1.3 & 0.8 & HARPS-pre \\
2454623.455 & -2.6 & 0.5 & HARPS-pre \\
2454684.921 & 5.7 & 0.3 & HARPS-pre \\
2454685.92 & 0.0 & 0.4 & HARPS-pre \\
2454712.027 & 0.7 & 4.9 & Lick \\
2454754.943 & 17.4 & 4.1 & Lick \\
2454777.943 & 11.0 & 4.0 & Lick \\
2454786.764 & 23.2 & 0.4 & HARPS-pre \\
2454788.865 & 28.0 & 0.3 & HARPS-pre \\
2454806.842 & 26.3 & 5.6 & Lick \\
2454866.0997 & 28.1 & 0.8 & UCLES \\
2454866.94 & 18.4 & 1.3 & UCLES \\
2454867.9158 & 26.6 & 1.3 & UCLES \\
2454869.0858 & 20.8 & 1.0 & UCLES \\
2454871.0348 & 30.4 & 1.3 & UCLES \\
2454882.81 & 26.1 & 6.3 & Lick \\
2454891.641 & 29.0 & 0.3 & HARPS-pre \\
2454911.667 & 21.4 & 6.1 & Lick \\
2455098.048 & -4.9 & 5.4 & Lick \\
2455121.051 & -15.3 & 6.1 & Lick \\
2455140.189 & -29.1 & 0.9 & UCLES \\
2455154.944 & -36.3 & 7.3 & Lick \\
2455174.901 & -21.7 & 5.3 & Lick \\
2455227.066 & -18.4 & 1.3 & UCLES \\
2455241.706 & 2.0 & 7.2 & Lick \\
2455278.689 & -14.0 & 5.7 & Lick \\
2455317.8583 & 0.8 & 0.6 & UCLES \\
2455447.026 & 19.3 & 5.7 & Lick \\
2455463.978 & 24.9 & 6.0 & Lick \\
2455525.2237 & 34.4 & 1.4 & UCLES \\
2455526.2103 & 38.8 & 0.9 & UCLES \\
2455566.876 & 25.3 & 4.8 & Lick \\
2455581.0932 & 43.3 & 1.1 & UCLES \\
2455588.839 & 28.6 & 4.8 & Lick \\
2455601.0 & 36.9 & 0.9 & UCLES \\
2455619.7 & 32.2 & 5.9 & Lick \\
2455649.7 & 18.5 & 4.6 & Lick \\
2455673.657 & -4.6 & 5.3 & Lick \\
2455706.843 & -5.3 & 1.0 & UCLES \\
2455783.3046 & -25.6 & 1.0 & UCLES \\
2455829.051 & -54.5 & 4.8 & Lick \\
2455831.046 & -53.2 & 5.0 & Lick \\
2455861.965 & -41.4 & 6.5 & Lick \\
2455879.2644 & -46.9 & 1.1 & UCLES \\
2455880.2195 & -38.9 & 0.9 & UCLES \\
2455892.848 & -63.2 & 5.7 & Lick \\
2455906.0446 & -38.4 & 1.1 & UCLES \\
2455969.9669 & -37.2 & 0.8 & UCLES \\
2455994.9599 & -20.4 & 0.9 & UCLES \\
2456051.8642 & 1.6 & 1.4 & UCLES \\
2456059.8647 & -4.0 & 1.6 & UCLES \\
2456301.679 & 47.8 & 0.3 & HARPS-pre \\
2456301.68 & 46.3 & 0.3 & HARPS-pre \\
2456316.618 & 49.2 & 0.3 & HARPS-pre \\
2456331.588 & 43.8 & 0.4 & HARPS-pre \\
2456343.512 & 42.3 & 0.3 & HARPS-pre \\
2456343.9918 & 47.7 & 0.9 & UCLES \\
2456346.511 & 47.3 & 0.5 & HARPS-pre \\
2456346.512 & 45.3 & 0.5 & HARPS-pre \\
2456361.487 & 48.0 & 0.3 & HARPS-pre \\
2456374.479 & 45.5 & 0.3 & HARPS-pre \\
2456374.882 & 55.7 & 1.2 & UCLES \\
2456377.9794 & 54.8 & 0.9 & UCLES \\
2456388.479 & 46.2 & 0.3 & HARPS-pre \\
2456404.47 & 44.8 & 0.3 & HARPS-pre \\
2456404.47 & 44.6 & 0.3 & HARPS-pre \\
2456526.2713 & -4.2 & 1.0 & UCLES \\
2456685.9759 & -27.9 & 0.9 & UCLES \\
2456747.9213 & -22.7 & 1.3 & UCLES \\
2457132.3415 & 2742.8 & 1.8 & SONG \\
2457136.3814 & 2752.8 & 1.4 & SONG \\
2457148.3413 & 2760.2 & 2.4 & SONG \\
2457282.707 & 2722.8 & 2.9 & SONG \\
2457324.5989 & 2713.4 & 3.0 & SONG \\
2457344.5478 & 2691.1 & 4.4 & SONG \\
2457390.5948 & 2688.4 & 2.6 & SONG \\
2457395.5167 & 2685.8 & 2.1 & SONG \\
2457398.3618 & 2688.4 & 4.3 & SONG \\
2457399.3624 & 2686.6 & 3.0 & SONG \\
2457406.5005 & 2690.0 & 2.4 & SONG \\
2457407.3848 & 2697.2 & 2.3 & SONG \\
2457414.4157 & 2699.8 & 3.0 & SONG \\
2457417.3935 & 2707.4 & 3.1 & SONG \\
2457420.4618 & 2694.4 & 6.4 & SONG \\
2457422.4698 & 2700.4 & 2.4 & SONG \\
2457424.438 & 2699.4 & 5.4 & SONG \\
2457428.4097 & 2706.4 & 3.5 & SONG \\
2457430.3314 & 2703.4 & 2.6 & SONG \\
2457454.3522 & 2692.9 & 1.9 & SONG \\
2457477.4264 & 2704.7 & 2.1 & SONG \\
2457497.3542 & 2712.8 & 2.4 & SONG \\
2457503.3745 & 2708.4 & 1.8 & SONG \\
2457504.3516 & 2713.8 & 1.7 & SONG \\
2457506.3559 & 2713.5 & 2.1 & SONG \\
2457507.3703 & 2706.1 & 2.0 & SONG \\
2457510.3447 & 2714.8 & 4.8 & SONG \\
2457683.632 & 2740.5 & 2.8 & SONG \\
2457684.634 & 2742.5 & 2.6 & SONG \\
2457696.6346 & 2745.0 & 3.0 & SONG \\
2457700.6429 & 2740.6 & 2.3 & SONG \\
2457701.6305 & 2731.4 & 2.3 & SONG \\
2457702.6354 & 2743.8 & 1.9 & SONG \\
2457713.5965 & 2736.9 & 2.4 & SONG \\
2457716.5983 & 2741.0 & 2.5 & SONG \\
2457717.6038 & 2740.2 & 2.1 & SONG \\
2457767.4856 & 2745.1 & 6.6 & SONG \\
2457768.4934 & 2759.3 & 3.3 & SONG \\
2457782.4046 & 2759.7 & 2.7 & SONG \\
2457818.355 & 2747.6 & 3.6 & SONG \\
2457863.3658 & 2719.8 & 1.9 & SONG \\
2458004.758 & 2709.1 & 2.8 & SONG \\
2458056.5996 & 2721.0 & 2.4 & SONG \\
2458070.7825 & 2699.1 & 2.3 & SONG \\
2458093.7678 & 2696.6 & 2.5 & SONG \\
2458106.6348 & 2694.7 & 2.5 & SONG \\
2458121.5482 & 2693.6 & 2.2 & SONG \\
2458183.3512 & 2701.7 & 1.9 & SONG \\
2458191.498 & -2.8 & 0.6 & HARPS-post \\
2458191.499 & -1.9 & 0.5 & HARPS-post \\
2458191.5 & -2.3 & 0.6 & HARPS-post \\
2458191.5 & -1.4 & 0.6 & HARPS-post \\
2458191.501 & -2.8 & 0.6 & HARPS-post \\
2458196.491 & -3.0 & 0.2 & HARPS-post \\
2458196.492 & -2.8 & 0.3 & HARPS-post \\
2458196.492 & -4.5 & 0.4 & HARPS-post \\
2458196.493 & -3.3 & 0.4 & HARPS-post \\
2458196.494 & -2.9 & 0.5 & HARPS-post \\
2458196.494 & -2.5 & 0.5 & HARPS-post \\
2458196.495 & -2.2 & 0.3 & HARPS-post \\
2458196.496 & -2.5 & 0.3 & HARPS-post \\
2458196.497 & -3.3 & 0.8 & HARPS-post \\
2458196.499 & -2.5 & 0.2 & HARPS-post \\
2458196.5 & -1.8 & 0.2 & HARPS-post \\
2458196.5 & -1.6 & 0.2 & HARPS-post \\
2458196.501 & -1.8 & 0.2 & HARPS-post \\
2458196.502 & -1.4 & 0.2 & HARPS-post \\
2458196.503 & -1.6 & 0.3 & HARPS-post \\
2458196.503 & -0.9 & 0.2 & HARPS-post \\
2458196.504 & -0.6 & 0.2 & HARPS-post \\
2458196.505 & -0.8 & 0.2 & HARPS-post \\
2458196.505 & -0.3 & 0.3 & HARPS-post \\
2458196.506 & -0.0 & 0.2 & HARPS-post \\
2458196.507 & 0.7 & 0.2 & HARPS-post \\
2458196.507 & 0.8 & 0.2 & HARPS-post \\
2458199.4683 & 2690.9 & 5.0 & SONG \\
2458221.47 & 8.3 & 0.2 & HARPS-post \\
2458221.471 & 7.8 & 0.2 & HARPS-post \\
2458221.471 & 7.7 & 0.2 & HARPS-post \\
2458221.472 & 7.9 & 0.3 & HARPS-post \\
2458221.473 & 7.2 & 0.3 & HARPS-post \\
2458221.473 & 7.8 & 0.3 & HARPS-post \\
2458221.474 & 7.7 & 0.3 & HARPS-post \\
2458221.475 & 6.9 & 0.3 & HARPS-post \\
2458221.476 & 6.8 & 0.3 & HARPS-post \\
2458221.476 & 7.4 & 0.3 & HARPS-post \\
2458224.3937 & 2705.0 & 3.4 & SONG \\
2458224.492 & 11.1 & 0.3 & HARPS-post \\
2458224.492 & 11.6 & 0.3 & HARPS-post \\
2458224.493 & 12.1 & 0.3 & HARPS-post \\
2458224.494 & 12.4 & 0.3 & HARPS-post \\
2458224.495 & 13.0 & 0.3 & HARPS-post \\
2458224.495 & 13.4 & 0.3 & HARPS-post \\
2458224.496 & 13.4 & 0.3 & HARPS-post \\
2458224.497 & 14.3 & 0.3 & HARPS-post \\
2458224.498 & 14.2 & 0.3 & HARPS-post \\
2458224.498 & 14.6 & 0.3 & HARPS-post \\
2458224.499 & 14.4 & 0.3 & HARPS-post \\
2458224.5 & 15.0 & 0.3 & HARPS-post \\
2458224.5 & 14.8 & 0.3 & HARPS-post \\
2458224.501 & 15.5 & 0.3 & HARPS-post \\
2458224.502 & 16.0 & 0.3 & HARPS-post \\
2458224.503 & 16.7 & 0.3 & HARPS-post \\
2458224.503 & 16.9 & 0.3 & HARPS-post \\
2458224.504 & 16.8 & 0.3 & HARPS-post \\
2458224.505 & 16.3 & 0.3 & HARPS-post \\
2458224.505 & 16.6 & 0.3 & HARPS-post \\
2458224.506 & 16.8 & 0.4 & HARPS-post \\
2458224.507 & 16.7 & 0.3 & HARPS-post \\
2458224.508 & 16.8 & 0.3 & HARPS-post \\
2458224.508 & 16.1 & 0.3 & HARPS-post \\
2458224.509 & 16.6 & 0.3 & HARPS-post \\
2458224.51 & 16.4 & 0.3 & HARPS-post \\
2458224.51 & 16.3 & 0.3 & HARPS-post \\
2458224.511 & 16.3 & 0.3 & HARPS-post \\
2458224.512 & 16.5 & 0.3 & HARPS-post \\
2458224.513 & 16.3 & 0.3 & HARPS-post \\
2458238.3509 & 2724.2 & 3.5 & SONG \\
2458350.7525 & 2753.8 & 3.7 & SONG \\
2458364.7141 & 2751.3 & 3.8 & SONG \\
2458378.7447 & 2754.8 & 3.4 & SONG \\
2458384.82 & 50.1 & 0.3 & HARPS-post \\
2458384.821 & 50.2 & 0.3 & HARPS-post \\
2458384.822 & 50.9 & 0.3 & HARPS-post \\
2458384.822 & 50.1 & 0.3 & HARPS-post \\
2458384.823 & 50.6 & 0.3 & HARPS-post \\
2458384.824 & 50.9 & 0.3 & HARPS-post \\
2458384.824 & 50.5 & 0.3 & HARPS-post \\
2458384.825 & 50.7 & 0.3 & HARPS-post \\
2458384.826 & 51.0 & 0.3 & HARPS-post \\
2458384.827 & 51.5 & 0.3 & HARPS-post \\
2458384.827 & 51.8 & 0.3 & HARPS-post \\
2458384.828 & 51.9 & 0.3 & HARPS-post \\
2458384.829 & 52.3 & 0.3 & HARPS-post \\
2458384.829 & 52.0 & 0.3 & HARPS-post \\
2458384.83 & 52.1 & 0.2 & HARPS-post \\
2458384.831 & 52.6 & 0.3 & HARPS-post \\
2458384.832 & 53.5 & 0.3 & HARPS-post \\
2458384.832 & 53.4 & 0.2 & HARPS-post \\
2458384.833 & 53.3 & 0.2 & HARPS-post \\
2458384.834 & 53.3 & 0.2 & HARPS-post \\
2458384.834 & 53.9 & 0.3 & HARPS-post \\
2458384.835 & 54.1 & 0.2 & HARPS-post \\
2458384.836 & 54.2 & 0.2 & HARPS-post \\
2458384.837 & 54.1 & 0.5 & HARPS-post \\
2458384.837 & 54.4 & 0.4 & HARPS-post \\
2458384.838 & 54.4 & 0.3 & HARPS-post \\
2458384.839 & 54.1 & 0.3 & HARPS-post \\
2458384.839 & 54.2 & 0.3 & HARPS-post \\
2458384.84 & 54.3 & 0.3 & HARPS-post \\
2458384.841 & 54.3 & 0.3 & HARPS-post \\
2458385.819 & 52.1 & 0.3 & HARPS-post \\
2458385.82 & 53.0 & 0.3 & HARPS-post \\
2458385.821 & 52.6 & 0.3 & HARPS-post \\
2458385.821 & 52.2 & 0.3 & HARPS-post \\
2458385.822 & 53.0 & 0.3 & HARPS-post \\
2458385.823 & 53.1 & 0.3 & HARPS-post \\
2458385.824 & 53.9 & 0.3 & HARPS-post \\
2458385.824 & 54.3 & 0.3 & HARPS-post \\
2458385.825 & 54.0 & 0.3 & HARPS-post \\
2458385.826 & 54.0 & 0.3 & HARPS-post \\
2458385.826 & 54.5 & 0.3 & HARPS-post \\
2458385.827 & 54.8 & 0.3 & HARPS-post \\
2458385.828 & 54.8 & 0.3 & HARPS-post \\
2458385.828 & 54.7 & 0.3 & HARPS-post \\
2458385.829 & 55.4 & 0.3 & HARPS-post \\
2458385.83 & 55.5 & 0.3 & HARPS-post \\
2458385.831 & 55.7 & 0.3 & HARPS-post \\
2458385.831 & 56.0 & 0.3 & HARPS-post \\
2458385.832 & 55.7 & 0.3 & HARPS-post \\
2458385.833 & 55.6 & 0.3 & HARPS-post \\
2458385.833 & 55.6 & 0.3 & HARPS-post \\
2458385.834 & 55.3 & 0.3 & HARPS-post \\
2458385.835 & 56.3 & 0.3 & HARPS-post \\
2458385.836 & 55.3 & 0.3 & HARPS-post \\
2458385.836 & 55.4 & 0.3 & HARPS-post \\
2458385.837 & 55.1 & 0.3 & HARPS-post \\
2458385.838 & 54.8 & 0.3 & HARPS-post \\
2458385.838 & 55.3 & 0.3 & HARPS-post \\
2458385.839 & 55.0 & 0.3 & HARPS-post \\
2458385.84 & 54.9 & 0.3 & HARPS-post \\
2458404.7743 & 2761.9 & 3.0 & SONG \\
2458421.699 & 2750.6 & 3.1 & SONG \\
2458439.5865 & 2755.5 & 3.7 & SONG \\
2458466.756 & 2744.4 & 4.3 & SONG \\
2458518.4756 & 2756.7 & 3.4 & SONG \\
2458533.3374 & 2745.2 & 3.1 & SONG \\
2458535.3333 & 2742.0 & 2.9 & SONG \\
2458545.4218 & 2750.0 & 3.2 & SONG \\
2458548.4526 & 2727.1 & 6.2 & SONG \\
2458566.4454 & 2738.1 & 3.6 & SONG \\
2458599.3684 & 2716.6 & 3.6 & SONG \\

        \noalign{\smallskip}
   \hline
   \end{longtable}
}

\end{appendix}

\end{document}